\documentclass[twocolumn,showpacs,preprintnumbers,amsmath,amssymb]{revtex4}

\usepackage{graphicx}
\usepackage{dcolumn}
\usepackage{bm}
\usepackage{float}

\begin{document}


\title{Multiple energy gap features of superconducting FeSe investigated by tunneling spectroscopy}

\author{Eike Venzmer}
\author{Alexander Kronenberg}
\author{Martin Jourdan}
 \email{Jourdan@uni-mainz.de}
\affiliation{Institut f\"ur Physik, Johannes Gutenberg-Universit\"at, Staudinger Weg 7, 55128 Mainz, Germany}

\date{\today}

\begin{abstract}

Planar tunneling junctions of the presumably unconventional superconductor FeSe were prepared and investigated. The junctions consisted of FeSe/AlO$_x$/Ag multilayers patterned lithographically into mesa structures. Bias voltage dependent differential conductivities dI/dV(V) of junctions in the tunneling, intermediate barrier strength, and metallic regimes were investigated. Depending on the barrier type between two and four features of the conductivity were obtained, which are discussed in the framework of multiple superconducting energy gaps. Specifically we reproduced with all barrier types an established energy gap and discovered an additional large gap structure. From two further conductivity features presumable one originates from another superconducting gap and the other from resonant states.

\end{abstract}

\maketitle

\section{Introduction}
Immediately after the discovery of the first iron based superconductor \cite{Kamihara2008}, many publications discussed possible unconventional pairing mechanisms \cite{Boeri2008, Haule2008, Mazin2008, Subedi2008}. These mechanisms are typically associated with specific symmetries of the order parameter. An extended s-wave state was proposed by Mazin \textit{et al.} \cite{Mazin2008}, which is supported by recent theoretical work \cite{Lischner2015}. Additionally, also d-wave states were predicted \cite{Kuroki2008}. However, only scarce direct experimental evidence for such symmetries is not available up to now.

Amongst the iron based superconductors FeSe has the simplest crystal structure with a critical temperature of $T_c \simeq 8.5\;\text{K}$ \cite{Hsu2008}. Critical temperatures up to $11.4\,\text{K}$ were obtained by biaxial strain resulting in an elongation of the c-axis \cite{Nabeshima2013} while a compression of this axis suppresses superconductivity \cite{Nie2009}. The critical temperature can be further increased up to $T_c \simeq 37\;\text{K}$ by applying hydrostatic pressure of $4.1\,\text{GPa}$ \cite{Medvedev2009}.

Theoretical descriptions of the electronic properties of FeSe are usually based on a five-band model with electron pockets at $(0,\pm \pi)$ and $(\pm \pi, 0)$ and hole pockets at (0,0) \cite{Gra09}. On this multiband Fermi surface different types of extended s-wave and d-wave order parameters are discussed, which include nodal as well as nodeless superconducting energy gaps. In-situ scanning tunneling spectroscopy (STS) on MBE grown FeSe thin films was interpreted as evidence for gap nodes by Song et al.\,\cite{Song2011a}. These authors observed one very clear single V-shaped energy gap with $\Delta_B \approx 2.2\,\text{meV}$ \cite{Song2011a, Song2011b, Song2012, Song2013}, which corresponds to ratios of $2\Delta_B/k_BT_c=5.5-6.4$. An additional conductivity feature at $2\Delta_C\approx9.4k_BT_c$, was interpreted as strong coupling to a spin fluctuation mode \cite{Song2014}. However, based on the multiple Fermi surfaces of FeSe, also multiple superconducting energy gaps are possible. Recent STS investigations of high quality FeSe single crystals by Kasahara \textit{et al.} observed additionally to a clear V-shaped energy gap with $\Delta_B=2.5\,\text{meV}$ ($2\Delta_B/k_BT_c=6.8$) a shoulder feature in the differential conductivity at $\simeq=3.5\,\text{meV}$, which they interpreted as a second superconducting gap with $2\Delta_C/k_BT_c=9.5$ \cite{Kasahara2014}. Also by Andre\`ev spectroscopy on break-junctions multiple features were observed and explained by two energy gaps with $\Delta_A\approx1.0\,\text{meV}$ ($2\Delta_A/k_BT_c=1.9$) and $\Delta_B\approx2.6\,\text{meV}$ ($2\Delta_B/k_BT_c=6.7$) \cite{Ponomarev2013}. Summarizing the available literature about the superconducting energy gap of FeSe agreement about a clear gap feature with $2\Delta_B/k_BT_c=5.5-6.7$ exists. However, the observation of additional conductivity features and their interpretation is inconsistent. 

Here we present spectroscopic measurements of the bias voltage dependent differential conductivity of planar FeSe/AlO$_x$/Ag tunneling junctions. We discuss three different kinds of spectra including tunneling like behavior, metallic point contact like behavior and data sets showing characteristics of both junction types.

\section{Junction preparation and measurement procedure}
Epitaxial FeSe thin films were prepared by rf-sputter deposition from two elementary 2~inch targets (Fe: Mateck GmbH, purity $99.99\%$; Se: Lesker, purity $99.999\%$) using AJA International A320-XP-MM sputtering sources. Superconducting thin films in (001)-orientation were obtained on MgO(100) substrates. For more information on the growth process and structural and morphological characterization of the $\beta$-FeSe thin films please see \cite{Venzmer2015}.

For junction preparation, on top of the $\beta$-FeSe thin films (thickness $d\approx 50\,\text{nm}$) thin Al layers ($2.8\,\text{nm}$) were sputter deposited and plasma oxidized for tunneling barrier formation. Sputter deposited Ag thin films ($d=50\,\text{nm}$) served as counter electrodes ($d=50\,\text{nm}$). Finally, the three layer samples were patterned into mesa structures of different sizes (from $60\times60\,\mu\text{m}^2$ to $100\times100\,\mu\text{m}^2$) by optical lithography and ion beam etching.

The bias voltage dependent differential conductivity of the junctions was measured by a standard a.c.-modulation technique (${\rm V_{ac}}=50\,\mu\text{V}$) in a $^3$He-cryostat. 

\section{Spectroscopic results}
None of our junctions showed a differential conductivity $dI/dV(V)$ approaching zero at zero bias, indicating that either the FeSe thin films were never homogeneously superconducting across the junction area or the AlO$_x$ thin layer never represented a perfect tunneling barrier. Figure 1 shows the differential conductivity of a junction, which is closest to tunneling dominated behavior as indicated by the largest ratio of the conductivities outside and inside the gap.

\begin{figure}[htb]
	\begin{center}
				\includegraphics[width=\columnwidth, angle=0]{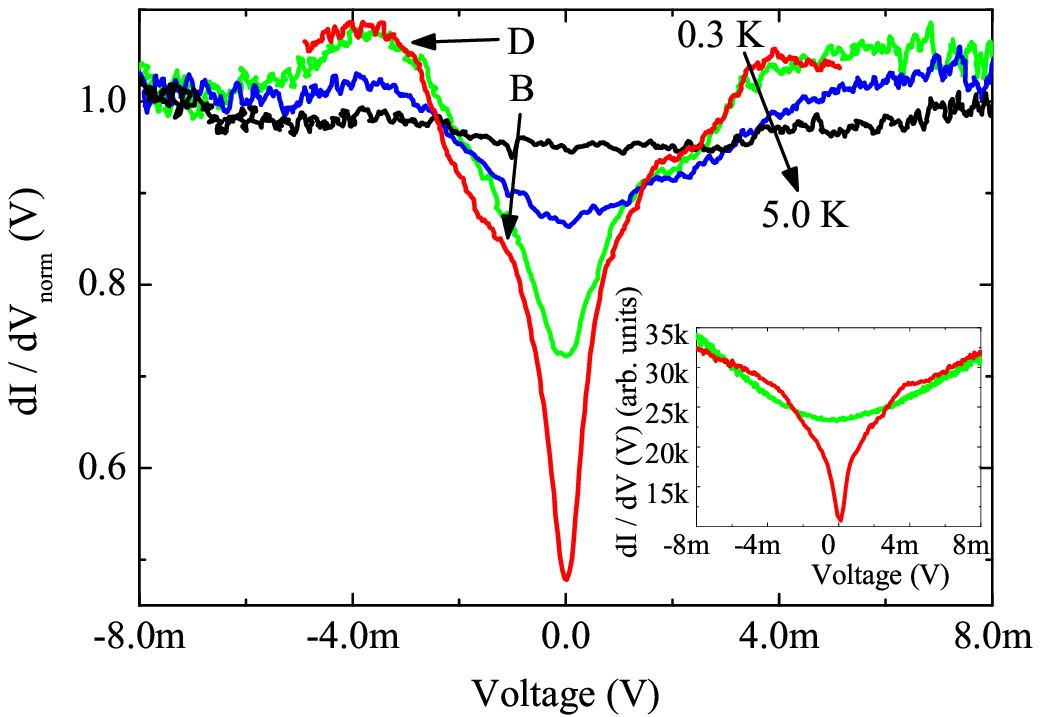}
				\caption{Differential conductivity $dI/dV_{norm}(V)$ of a FeSe/AlO$_x$/Ag junction at $T=0.3\,\text{K}$, 2K, 4K, and 5K, normalized to $dI/dV(V)$ measured at $T=6\,\text{K}$. The inset shows the conductivity at $T=0.3\,\text{K}$ and 6K without normalization.}
				\end{center}
				\label{fig:Tunnelspektren02}
\end{figure}		

The inset of Figure 1 shows the junction conductivities obtained at $T=0.3\,\text{K}$ and at the temperature $T=6\,\text{K}$ ($T_c\approx5\,\text{K}$) to which the conductivity curves displayed in the main panel were normalized.
The normalized spectrum measured at $T=0.3\,\text{K}$ shows a V-shaped gap feature in an energy range of $\pm 1.2\,\text{meV}$ around zero bias. Additionally, a second gap feature with conductivity peaks at $\pm 3.7\,\text{meV}$ is obvious.
Identifying both features with superconducting energy gaps $2\Delta_B/k_BT_c=5.6$ and $2\Delta_D/k_BT_c=17.2$ are obtained.


\begin{figure}[htb]
	\begin{center}
				\includegraphics[width=\columnwidth, angle=0]{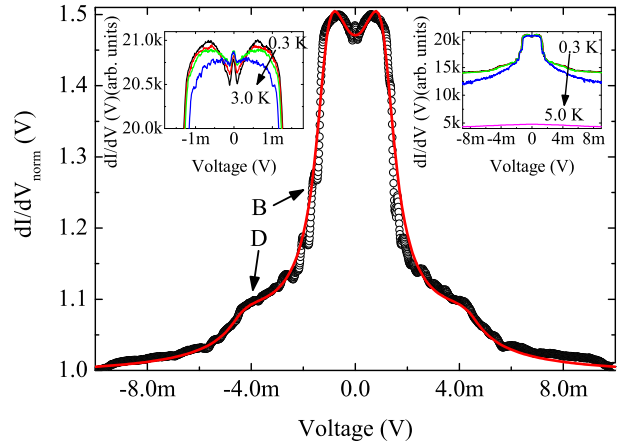}
				\caption{Differential conductivity $dI/dV_{norm}(V)$ of a FeSe/AlO$_x$/Ag junction at $T=0.3\,\text{K}$, normalized to $dI/dV(V)$ measured at $T=6\,\text{K}$ and BTK-fit assuming two energy gaps (Z = 0.3). The right inset shows the conductivities at $T=0.3\,\text{K}$, 3K and 5K without normalization. The left inset shows the differential conductivities at $T=0.3\,\text{K}$, 2K, and 3K.}
	\end{center}
	\label{fig:Andreevspektren04}
\end{figure}

A tunneling barrier with pinholes can result in a large variety of structures in the differential conductivity. These structures are often caused by local heating effects. However, small pinholes can behave analogous to point contacts representing metallic contacts with ballistic electron injections resulting in Andre\`ev reflection as described by the BTK-theory \cite{Blonder1982}. In Figure 2 a normalized Andre\`ev-like spectrum obtained measuring a FeSe/AlO$_x$/Ag junction is presented for temperatures between $0.3\,\text{K}$ and $5\,\text{K}$. The critical temperature of this sample amounts to $T_c=5\pm0.5\,\text{K}$, as indicated by an almost constant $dI/dV(V)$ curve measured at this temperature. As shown in the right inset of Figure 2, the $dI/dV(V)$ curve obtained just above $T_c$ is shifted to lower conductivities compared to the curves measured below $T_c$. This indicates that in the normal state of the FeSe base electrode, the pinhole(s) and the FeSe bulk resistivity within the mesa structure both contribute about half to the total mesa conductivity. Similar effects were observed in other ironpnictide superconductor-normal metal-superconductor junctions \cite{Doring2014}. Another deviation of the junction properties from the conventionally expected behavior of ballistic point contacts is the almost missing temperature dependence of the experimental differential conductivity up to $T_c$. The very weak temperature dependence observed is displayed in the left inset of figure 2. However, following the procedure of \cite{Daghero2010}, the experimental $dI/dV(V)$ curve measured at $0.3\,\text{K}$ can be fitted very well using the BTK-theory with Dynes broadening as shown in the main panel of Figure 2. The magnitudes of the gaps used in the fit are $\Delta_{B}=1.2\pm0.2\,\text{meV}$ and $\Delta_{D}=4.5\pm0.2\,\text{meV}$ corresponding to ratios of $2\Delta_{B}/k_BT_c=5.6\pm0.9$ and $2\Delta_{D}/k_BT_c=20.9\pm0.9$. 

The third type of spectra obtained investigating FeSe/AlO$_x$/Ag junctions, shows characteristics resembling both tunneling and Andre\`ev reflection, which is typical for junctions with intermediate barrier strength \cite{Blonder1982}. Compared to the junction with the Andre\`ev-like spectrum discussed above, only a small contribution ($\approx5\%$) of the bulk resistivity of the FeSe to the total measured mesa structure conductivity is deduced from the shift to lower conductivities in the normal state. In Figure 3 the temperature dependent conductivity of a junction with several features, which could be associated with superconducting energy gaps, is shown. 

\begin{figure}[htb]
	\begin{center}
				\includegraphics[width=\columnwidth, angle=0]{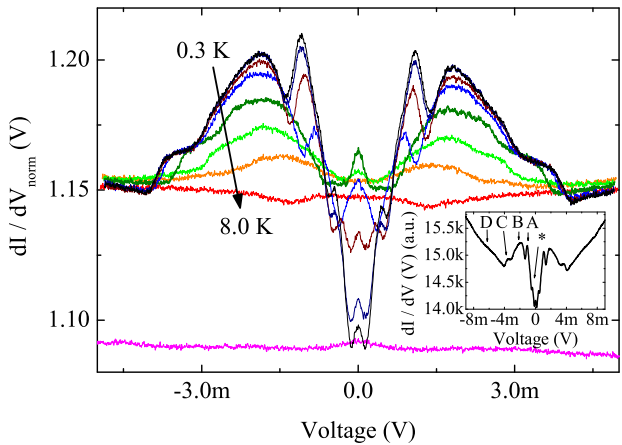}
				\caption{Differential conductivity $dI/dV_{norm}(V)$ of a FeSe/AlO$_x$/Ag junction at $T=0.3\,\text{K}$, 1K,2K, ... 8K, normalized to $dI/dV(V)$ measured at $T=9\,\text{K}$. The inset shows the conductivity at $T=1.3\,\text{K}$ without normalization, labeling different dI/dV(V) features.}
	\end{center}
	\label{fig:Andreevtunnelspektren01}
\end{figure}

The broad hump feature B at $\pm 1.9\,\text{meV}$ can be associated with the energy gap $\Delta_{B}$ discussed above (Figs.\,1 and 2), corresponding to $2\Delta_{B}/k_BT_c=5.6\pm1.2$.
The broad energy gap $\Delta_{D}$ discussed above corresponds to a kink labeled D in the dI/dV(V) curve at $\pm 7.5\,\text{meV}$ ($2\Delta_{D}/k_BT_c=21.5\pm2.9$).  Additionally, there are four more features of the differential conductivity, from which the very sharp feature A appears at $\pm 1.1\,\text{meV}$ ($2\Gamma_{A}/k_BT_c=3.2\pm0.6$). At higher bias voltages a clear shoulder at $\approx 3.7\,\text{meV}$ (labeled C) is observed, corresponding to $2\Delta_{C}/k_BT_c=10.6\pm1.2\,\text{meV}$. Further features are visible at $\pm 0.4\,\text{meV}$ and zero bias, which, based on the study of Fe atoms on a FeSe surface by Song et al.\,\cite{Song2011a}, can be explained as resonance states.

Please note that the features discussed above are reproduced by several junctions, which are in the intermediate barrier strength regime. The differential conductivity of another junction in the same regime is shown in Figure 5. Although the shape of the features discussed above is partially different and the critical temperature is strongly reduced to 4~K, all potential gaps can be found again at the same $2\Delta/k_BT_c$ values. 

\begin{figure}[htb]
	\begin{center}
				\includegraphics[width=\columnwidth, angle=0]{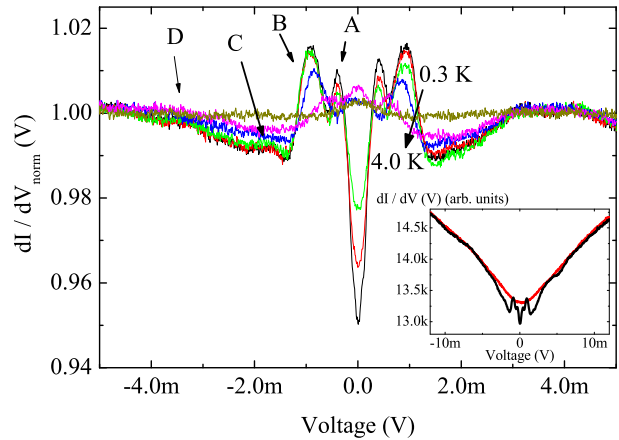}
				\caption{Differential conductivity $dI/dV_{norm}(V)$ of a FeSe/AlO$_x$/Ag tunneling junction at $T=0.3\,\text{K}$, 2K, 3K, and 4K, normalized to $dI/dV(V)$ measured at $T=5\,\text{K}$. The inset shows the conductivity at $T=1.4\,\text{K}$ and 5K without normalization.}
	\end{center}
	\label{fig:Andreevtunnelspektren02}
\end{figure}

The features A-D appear at bias voltages of $0.3\pm0.1\,\text{meV}$, $0.9\pm0.1\,\text{meV}$, $3.4\pm0.4\,\text{meV}$, and $5.5\pm1.0\,\text{meV}$, corresponding to $2\Gamma_A/k_BT_c=1.9\pm0.6$, $2\Delta_B/k_BT_c=5.4\pm0.6$, $2\Delta_C/k_BT_c=9.9\pm2.3$, and $2\Delta_D/k_BT_c=16.0\pm5.8$. 

\section{Discussion}
In the differential conductivity of all junctions two to four features can be associated with superconducting energy gaps. As such, they should scale with the critical temperature, which allows the comparison of samples with different quality reflected by different T$_c$ values. Thus in Table 1 an overview of the $2\Delta/k_BT_c$ values is given and compared with previously published data.

\begin{table*}[htb]
	\centering
		\begin{tabular}{l|cccc}
			from Figure & $2\Gamma_A/k_BT_c$ & $2\Delta_B/k_BT_c$ & $2\Delta_C/k_BT_c$ & $2\Delta_D/k_BT_c$\\
			\hline
			1 & - & $5.6\pm0.9$ & - & $17.2\pm1.9$\\
			2 & - & $5.6\pm0.9$ & - & $20.9\pm0.9$\\
			3 & $3.2\pm0.6$& $5.5\pm1.2$ & $10.6\pm1.2$ & $21.5\pm2.9$\\
		  5 & $1.9\pm0.6$ & $5.4\pm0.6$ & $10.4\pm1.2$ & $19.7\pm2.3$ \\
			\hline
			Kasahara \textit{et al.} \cite{Kasahara2014} & - & $6,8$ & $9,5$ & - \\
			Song \textit{et al.} \cite{Song2011a, Song2014} & - & $5,5-6,4$ & 9,4 & - \\
			Ponomarev \textit{et al.} \cite{Ponomarev2013} & $1,9$ & $6,7$ & - & - \\
		\end{tabular}
	\caption{Overview of $2\Delta/k_BT_c$ for all potential energy gaps / conductivity features.}
	\label{tab:Energielucken03}
\end{table*}

It is obvious that in all data sets an energy gap with $2\Delta_B/k_BT_c = 5.4-6.8$ is observed. Additionally, also the shape of the dI/dV(V) curves confirm that feature B is a clear superconducting energy gap.

Concerning the other gap like features none of them is observed simultaneously in all data sets.
We observed a sharp conductivity feature at $2\Gamma_A/k_BT_c = 1.3-3.8$, an energy range in which the existence of a superconducting energy gap was proposed based on break-junction spectroscopy by Ponomarev \textit{et al.} \cite{Ponomarev2013}. However, the features A in the differential conductivity spectra of our junctions with intermediate barrier strength are very sharp and more strongly suppressed by increasing temperature compared to the features B. In this sense they resemble resonance states as already discussed above for the lowest energy and zero bias conductivity peaks shown in Fig.\,3. 

Our data sets support the existence of a conductivity feature at $2\Delta_C/k_BT_c = 9.2-11.8$, which was proposed by Kasahara\textit{et al.} based on STS  to originate from a superconducting energy gap \cite{Kasahara2014}. However, considering the shape of the dI/dV(V) curves (Figs. 3 and 5) this could also be strong coupling feature as proposed by Song et al.\,\cite{Song2014}. 

Most important, we measured in all of our junctions an additional gap feature at relatively high energies with $2\Delta_D/k_BT_c = 15.3-24.3$, which cannot be found in the existing STS or break-junction spectroscopy data of FeSe. However, this feature appeared clearly in the most tunneling like (Fig.\,1) and in the most metallic (Fig.\,2) junction and shows in both cases the characteristics expected for a superconducting gap. 

As a simple test of the identification of the conductivity features B and C with superconducting energy gaps their temperature dependence is plotted in Figure 5. 

\begin{figure}[htb]
	\begin{center}
				\includegraphics[width=\columnwidth, angle=0]{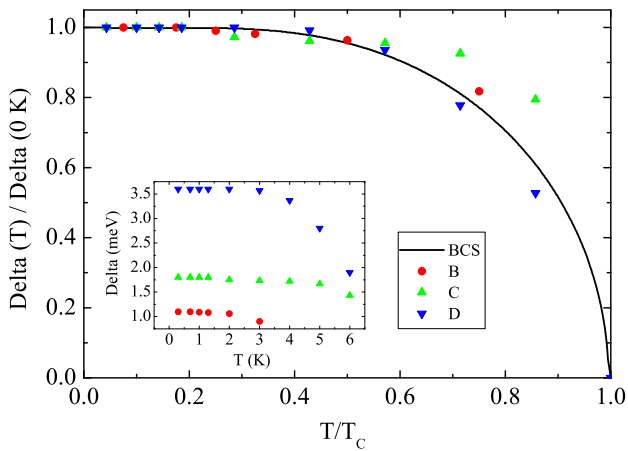}
				\caption{Temperature dependence of the normalized energy gaps, $\Delta_B$ and $\Delta_C$ as well as of the feature $\Gamma_A$ from Figure 3. Additionally, the temperature dependence $\Delta_{BCS}(T)/\Delta (0)$ as given by the BCS theory is plotted. The inset shows $\Delta_B(T)$, $\Delta_C(T)$, and $\Gamma_A(T)$ without normalization.}
	\end{center}
\end{figure}

As well $\Delta_B(T)$ as $\Delta_C(T)$, for which the association with an energy gap or a strong coupling effect is ambiguous, follow the prediction of the BCS theory supporting their identification as energy gaps. Additionally, also the temperature dependence of the resonance feature A is plotted, which disappears already well below $T_c$.   

\section{Summary}

The superconducting state of FeSe was investigated by tunneling spectroscopy on planar FeSe/AlO$_x$/Ag junctions. By an evaluation of the bias voltage dependent differential conductivities of junctions in the tunneling, intermediate barrier strength, and metallic regimes common features could be identified and associated with the superconducting density of states. Specifically, four characteristic features were observed. Whereas two of them show clear properties of superconducting gaps with $2\Delta_B/k_BT_c = 5.4-6.8$ and $2\Delta_D/k_BT_c = 15.3-24.3$, the interpretation is less clear for the other two. There are indications of a strong coupling effect with a boson excitation with $\omega_{boson}\approx 1.8$meV as previously proposed by Song et al. \cite{Song2014}. The fourth feature with $2\Gamma_A/k_BT_c = 1.3-3.8$ is presumably an impurity induced resonance state in the superconducting energy gap.\\

Financial support by the DFG (Jo404/6-1) is acknowledged. 
            
\newpage

\end{document}